\documentclass[amsmath,amssymb,pra,48pt,twocolumn,showpacs]{revtex4}
\usepackage{graphicx}
\usepackage{braket}
\usepackage{amsmath}
\usepackage{color}
\makeatletter
\renewcommand*\env@matrix[1][\arraystretch]{%
  \edef\arraystretch{#1}%
  \hskip -\arraycolsep
  \let\@ifnextchar\new@ifnextchar
  \array{*\c@MaxMatrixCols c}}
\makeatother
\numberwithin{equation}{section}

\begin{document}

\title{Three-photon polarization ququarts: polarization, entanglement and Schmidt decompositions}
\author{M.V. Fedorov}
 \email{fedorovmv@gmail.com}
\affiliation{A.M. Prokhorov General Physics Institute, Moscow, Russia}
\author{N.I. Miklin}
\email{miklinnikolai@gmail.com}
\affiliation{Moscow Institute of Physics and Technology, Dolgoprudny, Moscow Region, Russia}
\date{\today}
\begin{abstract}
We consider polarization states of three photons, each in the same given spectral-angular mode. A general form of such states is a superposition of four basic three-photon polarization modes, to be referred to as three-photon polarization ququarts. All such states can be considered as consisting of one- and two-photon parts, which can be entangled with each other. The degrees of entanglement and polarization as well as the Schmidt decomposition and Stokes vectors of three-photon polarization ququarts are found and discussed.
\end{abstract}


\maketitle

\numberwithin{equation}{section}
\section{Introduction}

The main objects of the modern science of quantum information are bipartite and, in particular, biphoton states. Characteristic features of such states are their entanglement and polarization, which are widely investigated and used in practical purposes, such as, e.g., transmission of information in quantum nets. The simplest biphoton states are purely polarization states of two photons belonging both to the same spatial and spectral mode (biphoton polarization qutrits). Biphoton states can be produced in different ways, but the most often used method is based on the phenomenon of Spontaneous Parametric Down-Conversion (SPDC) in nonlinear birefringent crystals. In such processes some of pump photons are converted in crystals into pairs of photons of smaller frequencies, and the pump is taken not too strong to avoid simultaneous production of four and higher amounts of photons. On the other hand, in stronger fields, multiphoton quantum states arise rather naturally in the process of parametric amplification \cite{Kok,Banaszek-03}, and the first reports on analysis of entanglement achievable in such ``macroscopic quantum states$"$ was given in the works \cite{Agafonov, Ishhakov}. Note, however, that definitions of the degree of entanglement in multiphoton states can be  not as simple as in the case of biphotons, and further investigations may be needed. In this paper we consider from this point of view the simplest quantum states more complicated than biphoton states, which are states of simultaneously produced three photons.

A general interest to three-photon quantum states exists since rather long ago and persists until nowadays \cite{Braunstein, Elutin, Banaszek, Mlynek, Keller, Boumeester, JWPan1, Acin, Dur1, JWPan2, Gisin, Hnilo,  Zeilinger, Che-Ivanova, Corona, Ding}. In principle, production of three-photon states can be realized in the usual SPDC scheme but with not-birefringent crystals having central symmetry. In this case the usual SPDC is forbidden as the second-order susceptibility equals zero, $\chi^{(2)}=0$, whereas the third-order susceptibility is nonzero, $\chi^{(3)}\neq 0$, and the three-photon decay of pump photons is possible. It's true that usually $\chi^{(3)}$ is very small, and to have efficient three-photon SPDC generation one has to use too strong pump fields. But in some semiconductor crystals (GaAs, Si, InSb) the third-order susceptibility can be rather high and comparable with typical second-order susceptibilities  birefringent crystals. Another possibility of making the 3rd-order processes efficient is related to the use of fibers \cite{Corona}, where a small value of the 3rd-order susceptibility can be compensated by a long distance at which the three-photon decay of pump photons can occur in fibers. Finally, one approach more is based on the use of double SPDC pairs containing four photons, with one of them subsequently set apart \cite{Zeilinger}.
$\quad\quad\quad$
\indent In this paper we consider theoretically the most general form of three-photon pure polarization states with collinearly propagating photons of and coinciding given frequencies. There are only four basic configurations of such states and for their superpositions we will use the name of Three-Photon Polarization Ququarts (TPPQ). In a general case, TPPQ can be considered as consisting of one- and two-photon parts. Owing to indistinguishability of photons such presentation and its features are unique for any given TPPQ and do not depend of which photons are selected to belong to one-photon and which to two-photon parts. The presentation of TPPQ states as consisting of one- and two-photon parts arises naturally in the description in terms of polarization wave functions depending on three discrete polarization variables of three photons. As shown below for any TPPQ its three-particle wave function can be presented in the form of the Schmidt decomposition \cite{Schmidt,Grobe,Knight,CP}, i.e. in the form of a sum of two products of single-photon and two-photon wave functions. Similar decompositions occur also for the two-photon and single-photon reduced density matrices of TPPQ, with reduction defined as taking traces of the total TPPQ density matrix over one or two photon variables. The reduced density matrices and their eigenvalues determine the degree of entanglement between one- and two-photon components of TPPQ, as well as the TPPQ Stokes vectors and degree of polarization. All these parameters are found below in a general form and analyzed in details in a series of the most representative examples. Some schemes for measuring TPPQ parameters in experiments are discussed.
\section{Three-photon polarization ququarts}
As defined above, TPPQ states are states of three photons with arbitrary distributed polarizations  but with all three photons belonging to the same single frequency-angular mode (in the simplest case, having the same identical given frequency and collinear wave vectors). In such cases photons have the only degree of freedom in which they can be entangled or not, and this is the polarization degree of freedom. There are only four tree-photon polarization modes $(3_H)$, $(2_H,1_V)$, $(1_H,2_V)$, $(3_V)$, where the numbers 1, 2, 3 indicate amounts of photons and, as usual, the labels $H$ and $V$ indicate horizontal and vertical polarizations (in the plane $(x,y)$ perpendicular to the direction of propagation of photons along the $z$-axis). These four three-photon modes correspond to the following four TPPQ basic state vectors:
\begin{gather}
 \nonumber
 \ket{3_H}=\frac{a_H^{\dag^{\,3}}}{\sqrt{6}}\ket{0},\;\ket{2_H,1_V}=\frac{a_H^{\dag^{\,2}}a_V^\dag}{\sqrt{2}}\ket{0},\\
 \label{Bas-st-vect}
 \;\ket{1_H,2_V}=\frac{a_H^\dag a_V^{\dag^{\,2}}}{\sqrt{2}}\ket{0}, \;\ket{3_V}=\frac{a_V^{\dag^{\,3}}}{\sqrt{6}}\ket{0}.
\end{gather}
A general TPPQ state is determined as a superposition of four basic state vectors of Eq. (\ref{Bas-st-vect})
\begin{equation}
 \nonumber
 \ket{\Psi}=C_1\ket{3_H}+C_2\ket{2_H,1_V}+C_3\ket{1_H,2_V}+C_4\ket{3_V}
\end{equation}
\small
\begin{equation}
 \label{qqrt-st-vect}
 =\left(\frac{C_1}{\sqrt{6}}a_H^{\dag\,3}+\frac{C_2}{\sqrt{2}}a_H^{\dag\,2}a_V^\dag+
 \frac{C_3}{\sqrt{2}}a_H^\dag a_V^{\dag\,2}+ \frac{C_4}{\sqrt{6}}a_V^{\dag\,3}\right)\ket{0},
\end{equation}
\normalsize
where $C_{1,2,3,4}$ are arbitrary complex constants restricted only by the normalization condition $\sum_i|C_i|^2=1$.  Also, as well known, the global phase of the superposition (\ref{qqrt-st-vect}) does not affect any possible measurements and can be taken having any given value. In particular, this global phase can be chosen to make one of the constants $C_i$ real, which will be used in some of our further derivations. Thus, as four complex constants $C_i$ are equivalent to eight real constants and as two of these eight constants can be discounted, an arbitrary three-photon polarization ququart is characterized completely by six real constants.

It should be emphasized that four configurations, four basic states, and four constants $C_i$ occur only owing to indistinguishability of photons. It's easy to imagine other cases of three distinguishable particles, one-qubit each. For example one can think about three different two-level atoms $a$, $b$, and $c$, but with identical ground and excited levels $E_g$ and $E_e$. In this case each of two configurations $(2_H,1_V)$ and $(1_H,2_V)$ turns into three different configurations
\begin{gather}
 \nonumber
 (2_H,1_V)\rightarrow (a_g;\,b_g;\,c_e),\;(a_g;\,b_e;\,c_g),\;(a_e;\,b_g;\,c_g),\\
 \nonumber
 (1_H,2_V)\rightarrow (a_g;\,b_e;\,c_e),\;(a_e;\,b_g;\,c_e),\;(a_e;\,b_e;\,c_g).
\end{gather}
This gives eight configurations totally, eight terms in the superposition substituting that of Eq. (\ref{qqrt-st-vect}), and eight complex constants $C_i$, or $2\times 8-2=14$ independent real constants. This is not the case we consider here. We consider only pure three-photon polarization states with indistinguishability of photons taken into account, which restricts by four the amounts of three-photon basic modes and of constants $C_i$.

The density matrix of the state $\ket{\Psi}$ (\ref{qqrt-st-vect}) is defined as $\hat{\rho}=\ket{\Psi}\bra{\Psi}$. Matrix elements of $\hat{\rho}$ are given by
\begin{gather}
 \nonumber
 \rho(\sigma_1,\sigma_2, \sigma_3;\sigma_1^\prime,\sigma_2^\prime,\sigma_3^\prime)=
 \braket{\sigma_1,\sigma_2, \sigma_3|\hat{\rho}|\sigma_1^\prime,\sigma_2^\prime,\sigma_3^\prime}\\
 \label{matr-el}
 =\Psi(\sigma_1,\sigma_2,\sigma_3)\Psi^*(\sigma_1^\prime,\sigma_2^\prime,\sigma_3^\prime),
\end{gather}
where $\sigma_1$, $\sigma_2$ and $\sigma_3$ are polarization variables of three indistinguishable photons, and $$\Psi(\sigma_1,\sigma_2,\sigma_3)=\braket{\sigma_1,\sigma_2,\sigma_3|\Psi}$$ is the wave function of TPPQ. Each of three polarization variables can take independently only two values, either $H$ or $V$. As photons are indistinguishable particles and as they are bosons, the wave function $\Psi(\sigma_1,\sigma_2,\sigma_3)$ must be symmetric with respect to all transpositions of variables $\sigma_1$, $\sigma_2$, and $\sigma_3$.

Both the general TPPQ wave function and the wave functions of basic states (\ref{Bas-st-vect}) can be expressed in terms of products of single-photon polarization wave functions
\begin{gather}
 \nonumber
 \psi_H(\sigma_i)=\braket{\sigma_i|1_H}=\delta_{\sigma_i,H}\equiv\left(1\atop 0\right)_i,\\
 \label{one-phot-wf}
 \psi_V(\sigma_i)=\braket{\sigma_i|1_V}=\delta_{\sigma_i,V}\equiv\left(0\atop 1\right)_i,
\end{gather}
where the upper and lower lines in columns correspond, respectively, to the horizontal and vertical polarizations. The labels $i=1,2,3$  in the formulas of Eq. (\ref{one-phot-wf}) numerate variables of three photons in three-photon states. Of course, this numeration of variables does not add to photons any additional degrees of freedom and does not make photons distinguishable. As said above and is well known, indistinguishability of photons results in the requirement of symmetry of multiphoton wave functions with respect to variable transpositions. Actually, even without these explanations general rules for finding multiphoton wave functions from the corresponding state vectors are well known in quantum electrodynamics \cite{Schweber}, and these rules include summation of products of one-photon wave functions over all transpositions. With these remarks taken into account, the wave functions of the basic states (\ref{Bas-st-vect}) can be written as
\begin{gather}
 \nonumber
 \Psi_{3_H}(\sigma_1,\sigma_2,\sigma_3)=\braket{\sigma_1,\sigma_2,\sigma_3|3_H}=
 \delta_{\sigma_1,H}\delta_{\sigma_2,H}\delta_{\sigma_3,H}\\
 \label{Psi-3H}
 =\left(1\atop 0\right)_1\left(1\atop 0\right)_2\left(1\atop 0\right)_3,
\end{gather}
\begin{gather}
 \nonumber
 \Psi_{2_H,1_V}(\sigma_1,\sigma_2,\sigma_3)=\braket{\sigma_1,\sigma_2,\sigma_3|2_H,1_V}=\\
 \nonumber
 \frac{\delta_{\sigma_1,H}\delta_{\sigma_2,H}\delta_{\sigma_3,V}
 +\delta_{\sigma_1,H}\delta_{\sigma_2,V}\delta_{\sigma_3,H}
 +\delta_{\sigma_1,V}\delta_{\sigma_2,H}\delta_{\sigma_3,H}}{\sqrt{3}}\\
 \nonumber
 =\frac{1}{\sqrt{3}}\Bigg\{\left(1\atop 0\right)_1\left(1\atop 0\right)_2\left(0\atop 1\right)_3+\left(1\atop 0\right)_1\left(0\atop 1\right)_2\left(1\atop 0\right)_3\\
 \label{Psi-2H-1V}
 +\left(0\atop 1\right)_1\left(1\atop 0\right)_2\left(1\atop 0\right)_3\Bigg\},
\end{gather}
\begin{gather}
 \nonumber
 \Psi_{1_H,2_V}(\sigma_1,\sigma_2,\sigma_3)=\braket{\sigma_1,\sigma_2,\sigma_3|1_H,2_V}=\\
 \nonumber
 \frac{\delta_{\sigma_1,H}\delta_{\sigma_2,V}\delta_{\sigma_3,V}
 +\delta_{\sigma_1,V}\delta_{\sigma_2,H}\delta_{\sigma_3,V}
 +\delta_{\sigma_1,V}\delta_{\sigma_2,V}\delta_{\sigma_3,H}}{\sqrt{3}}\\
 \nonumber
 =\frac{1}{\sqrt{3}}\Bigg\{\left(1\atop 0\right)_1\left(0\atop 1\right)_2\left(0\atop 1\right)_3+\left(0\atop 1\right)_1\left(1\atop 0\right)_2\left(0\atop 1\right)_3\\
 \label{Psi-1H-2V}
 +\left(0\atop 1\right)_1\left(0\atop 1\right)_2\left(1\atop 0\right)_3\Bigg\},
\end{gather}
\begin{gather}
 \nonumber
 \Psi_{3_V}(\sigma_1,\sigma_2,\sigma_3)=\braket{\sigma_1,\sigma_2,\sigma_3|3_V}=
 \delta_{\sigma_1,V}\delta_{\sigma_2,V}\delta_{\sigma_3,V}\\
 \label{Psi-3V}
 =\left(0\atop 1\right)_1\left(0\atop 1\right)_2\left(0\atop 1\right)_3.
\end{gather}
For shortening formulas we have dropped the direct-product symbols $\otimes$ between columns in Eqs. (\ref{Psi-3H})-(\ref{Psi-3V}). Note that the forms of writing three-photon wave functions via Kroneker symbols and via products of two-line columns are absolutely equivalent, and they are reproduced here together only for emphasizing this equivalence. Note also that often single-photon polarization wave functions are written in the Dirac form $\ket{H}$ and $\ket{V}$, which makes them indistinguishable from the state vectors. In application to multiphoton states, the use of Dirac notations for single-photon wave functions requires using indices $i$ for indication of variables on which these functions depend. Then Eqs. (\ref{Psi-3H})-(\ref{Psi-3V}) can be rewritten in the same form but with the substitution of $\delta_{\sigma_i,H}\equiv\left(1\atop 0\right)_i$ by $\ket{H}_i$ and $\delta_{\sigma_i,V}\equiv\left(0\atop 1\right)_i$ by $\ket{V}_i$. But, inevitably,  for making any further transformations or manipulations with the wave functions one has to return either to the Kroneker-symbol or to the matrix forms of Eqs. (\ref{Psi-3H})-(\ref{Psi-3V}).

Superposition of the basic three-photon polarization wave functions (\ref{Psi-3H})-(\ref{Psi-3V}) with the same coefficients as in Eq. (\ref{qqrt-st-vect}) gives the wave function of TPPQ in a general form
\begin{gather}
 \nonumber
 \Psi(\sigma_1,\sigma_2,\sigma_3)=C_1\Psi_{3_H}(\sigma_1,\sigma_2,\sigma_3)+C_2\Psi_{2_H,1_V}(\sigma_1,\sigma_2,\sigma_3)\\
 \label{qqrt-wf-gen}
 +C_3\Psi_{1_H,2_V}(\sigma_1,\sigma_2,\sigma_3)+C_4\Psi_{3_V}(\sigma_1,\sigma_2,\sigma_3).
\end{gather}

\section{Eigenvalues of the reduced density matrices and the degrees of polarization and entanglement}

The density matrix $\rho$ (\ref{matr-el}) can be reduced, e.g., at first, with respect to the variable $\sigma_3$ to give the reduced two-photon density matrix $\rho_r^{(1,2)}(\sigma_1,\sigma_2;\sigma_1^\prime,\sigma_2^\prime)$. Then this matrix can be further reduced with respect to the variable $\sigma_2$ to give the twice reduced single-photon density matrix $\rho_{rr}^{(1)}(\sigma_1;\sigma_1^\prime)$. In the matrix form $\rho_{rr}$ is given by
\begin{gather}
 \nonumber
 \rho_{rr}^{(1)}=Tr_{2}\rho_r^{(1,2)}=Tr_{2,3}\rho=\\
  \label{rho-rr}
 \scriptsize
 \left(
 \begin{matrix}
  |C_1|^2+\dfrac{2|C_2|^2}{3}+\dfrac{|C_3|^2}{3} & \dfrac{C_1C^*_2}{\sqrt{3}} + \dfrac{2C_2C^*_3}{3} + \dfrac{C_3C^*_4}{\sqrt{3}}\\
  \dfrac{C^*_1C_2}{\sqrt{3}} + \dfrac{2C^*_2C_3}{3} + \dfrac{C^*_3C_4}{\sqrt{3}} & \dfrac{|C_2|^2}{3} + \dfrac{2|C_3|^2}{3} + |C_4|^2
 \end{matrix}\right)
\end{gather}
Owing to symmetry of the wave function (\ref{qqrt-wf-gen}) with respect to variable transpositions, the twice reduced density matrix $\rho_{rr}$ is unique for any given three-photon state and does not depend of a choice of variables with respect to which the total density matrix $\rho$ is reduced, ($2,3$), or ($1,2$), or ($1,3$). The same is true for the two-photon reduced density matrix $\rho_r$: its form does not depend of a  choice of a single variable with respect to the which the total density matrix $\rho$ is reduced to give $\rho_r$.

The $2\times 2$ twice-reduced density matrix $\rho_{rr}$ (\ref{rho-rr}) can be diagonalized, and its eigenvalues $\lambda$ can be shown to obey the equation
\begin{equation}
 \label{lambda-eq}
 \lambda^2 - \lambda + \frac{1}{4}C_g^2 = 0,
\end{equation}
which has two solutions
\begin{equation}
  \label{lambda-pm}
  \lambda_\pm=\frac{1}{2}\bigg(1\pm\sqrt{1-C_g^{\,2}}\,\bigg),
 \end{equation}
 obeying the normalization condition $\lambda_++\lambda_-=1$.

 $C_g$ in Eqs. (\ref{lambda-eq}) and (\ref{lambda-pm}) is the parameter, which can be interpreted as the generalized concurrence, and for which we find  the expressions
\begin{gather}
    \nonumber
    C_g =2\sqrt{\lambda_+\lambda_-}=2\sqrt{\lambda_-(1-\lambda_-)}=2\Bigg[2\left|\frac{C_1C_3}{\sqrt{3}} - \frac{C^2_2}{3}\right|^2\\
    \label{Conc}
     + \left|C_1C_4 - \frac{C_2C_3}{3}\right|^2+ 2\left|\frac{C_4C_2}{\sqrt{3}} - \frac{C^2_3}{3}\right|^2\Bigg]^{1/2}.
\end{gather}

\begin{figure}[h]
\includegraphics[width=6cm]{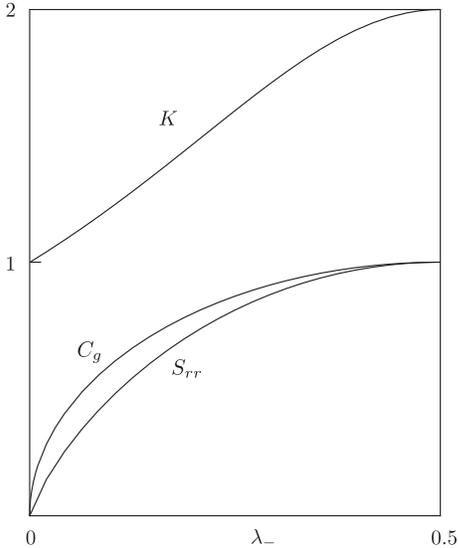}
\caption{{\protect\footnotesize {Generalized concurrence $C_g$ (\ref{Conc}),von Neumann entropy of the twice reduced density matrix $S_{rr}$ (\ref{entr}) and the Schmidt parameter $K$ (\ref{K}) as functions of the parameter $\lambda_-$ (\ref{lambda-pm})}}}
\label{Fig1}
\end{figure}

The matrix, adjoint to $\rho_{rr}^{(1)}$, is the two-photon reduced density matrix $\rho_{r}^{(2,3)}=Tr_1\rho$. This matrix has the dimensionality $4\times 4$, but, as it should be \cite{Peres},  it has only two non-zero eigenvalues coinciding with $\lambda_\pm$ of Eq. (\ref{lambda-pm}). Though the general expression (\ref{Conc}) for the parameter $C_g$ in terms of constants $C_i$ is much more complicated than the corresponding expression for concurrence of biphoton polarization states (qutrits) \cite{NJP}, the relations between the generalized concurrence  and eigenvalues $\lambda_\pm$ of $\rho_{rr}$ are identical to those occurring in the biphoton case. The same is true for many further relations between parameters characterizing the degrees of entanglement and polarization to be derived below. Note, however, that the generalized concurrence $C_g$ is not exactly the same as the concurrence introduced by C. K.  Wootters for two-qubit bipartite states \cite{Wootters}. To remind, for pure bipartite states the Wootters' concurrence can be defined as  $C_W=|\braket{\Psi|\widetilde{\Psi}}|$, where $\widetilde{\Psi}$ is the spin-flipped complex conjugate wave function, $\widetilde{\Psi} =\prod_i (\sigma_y)_i\Psi^*$, and $(\sigma_y)_i$ are the $y$-Pauli matrices for all $i$-th polarization variables. It can be easily found that  for TPPQ (\ref{qqrt-st-vect}), (\ref{qqrt-wf-gen}) this definition gives $C_W\equiv 0$, whereas the generalized concurrence $C_g$ (\ref{Conc}) can take any values in the interval $[0,1]$ depending on values of the constants $C_i$. We assume that for TPPQ the generalized concurrence is a good entanglement quantifier. This assumption is supported, in particular, by perfect compatibility of the generalized concurrence $C_g$ with such another entanglement quantifier as the von Neumann entropy of the double reduced density matrix
\begin{gather}
 \nonumber
 S_{rr}=-\lambda_+\log_2\lambda_+-\lambda_-\log_2\lambda_-\\
 \label{entr}
 =-\lambda_-\log_2\lambda_- -(1-\lambda_-)\log_2(1-\lambda_-).
\end{gather}
In Fig. \ref{Fig1} the functions $S_{rr}(\lambda_-)$ and $C_g(\lambda_-)$ are plotted together.
Both of them show that, as mentioned above, the
degree of entanglement of three-photon states varies from 0 to 1. An example of maximally entangled state (with $S_{rr}=C_g(\lambda_-)=1$) is $\ket{\Psi}=\frac{1}{\sqrt{2}}(\ket{3_H}+\ket{3_V})$, with $\lambda_+=\lambda_-=\frac{1}{2}$.The opposite case of a disentangled state is the state with $C_1=1$,  $C_{2,3,4}=0$,  $\ket{\Psi}=\ket{3_H}$, $\lambda_+=1$, $\lambda_-=0$, and $S_{rr}=C_g=0$.
The third curve  shown in Fig. \ref{Fig1} is the Schmidt entanglement parameter
\begin{equation}
 \label{K}
 K=Tr\rho_{rr}^2=\frac{1}{\lambda_+^2+\lambda_-^2}=\frac{2}{2-C_g^2}.
\end{equation}
As a function of $\lambda_-$, the Schmidt parameter $K(\lambda_-)$ grows monotonously and synchronously with $C_g(\lambda_-)$ and $S_{rr}(\lambda_-)$ from $K(0)=1$ to $K(0.5)=2$. As often said, all three parameters, $C_g$, $S_{rr}$ and $K$, characterize the same degree of entanglement of  three-photon states, though in different metrics.

Degree of polarization is a characteristics of both biphoton and three-photon quantum states, complementary to their degree of entanglement. Mathematically the degree of polarization per one photon is defined as $P=\left|\vec{S}\right|$ where $\vec{S}$ is the one-photon Stokes vector
\begin{equation}
 \label{Stokes-def}
 \vec{S} = Tr(\rho_{rr}\vec{\sigma}),
\end{equation}
and $\vec{\sigma}$ is the vector of Pauli matrices. Eq. (\ref{Stokes-def}) means that the twice reduced density matrix $\rho_{rr}$ is equivalent to the polarization matrix \cite{Che}
\begin{equation}
 \label{pol-matr}
 \rho_{rr}=\rho_{pol}=\frac{1}{2}
 \left(\begin{matrix}
 1+S_3 & S_1-iS_2\\
 S_1+iS_2 & 1-S_3
 \end{matrix}\right).
\end{equation}
Numeration of axes is related to their orientation in the Poincar\'{e} sphere (see Fig. \ref{Fig4} below): the numbers 3, 1 and 2 correspond, respectively, to the horizonal-vertical, $(-45^\circ,45^\circ)$, and left-right circular polarization axes. By comparing Eqs. (\ref{pol-matr}) and (\ref{rho-rr}), we easily find the Stokes vectors of TPPQ in a general form
\begin{gather}
 \vec{S}=
 \label{Stokes-gen}
 \left(
 \begin{matrix}
  2{\rm Re}\left(\displaystyle\frac{C_1C^*_2+C_3C^*_4}{\sqrt{3}}+\frac{2C_2C^*_3}{3}\right)
  \\
  -2{\rm Im}\left(\displaystyle\frac{C_1C^*_2+C_3C^*_4}{\sqrt{3}}+\frac{2C_2C^*_3}{3}\right)\\
  |C_1|^2+\dfrac{|C_2|^2}{3} - \dfrac{|C_3|^2}{3}-|C_4|^2 \\
 \end{matrix}\right).
\end{gather}
With a simple algebra, it can be shown that for TPPQ the relations between the degree of polarization $P$ and the generalized concurrence $C_g$ and Schmidt parameter $K$ remain the same as earlier derived relations between the degrees of polarization and entanglement of biphoton qutrits \cite{NJP}:
\begin{equation}
 \label{P-C-K}
 P^2 + C_g^2 =P^2 + 2\left(1-K^{-1}\right)= 1.
\end{equation}
Note that the total Stokes vector of a three-photon state is three times longer and the degree of polarization is three times higher than the single-photon Stokes vector and the  degree of polarization per one photon, ${\vec S}^{\,tot}=3{\vec S}^{single}$ and  $P^{tot}=3P^{single}$.

\section{Special cases}

In two examples to be considered in more details are those with only two non-zero terms in a general definition of  three-photon ququatrts of Eq. (\ref{qqrt-st-vect}).

\subsection{$C_2=C_3=0$}
\begin{figure}[h]
\includegraphics[width=6cm]{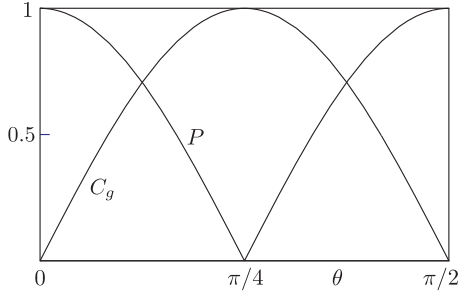}
\caption{{\protect\footnotesize {Degrees of entanglement $C_g$ and polarization $P$ (per one photon) as functions of the parameter $\theta$ of Eq. (\ref{case-a})}}}\label{Fig2}
\end{figure}

The state vector (\ref{qqrt-st-vect}) takes the form
\begin{gather}
 \nonumber
 \ket{\Psi}=C_1\ket{3_H}+C_4\ket{3_V}\\
 \label{case-a}
 =\cos\theta\ket{3_H}+e^{i\varphi}\sin\theta\ket{3_V},
\end{gather}
where in the parametrization with two real constants $\theta$ and $\phi$ the phase of $C_1$ is taken equal zero, $\pi\geq\theta\geq 0$ and $\pi/2\geq\varphi\geq-\pi/2$. In this case the general equations (\ref{lambda-pm}), (\ref{Conc}), and (\ref{P-C-K}) yield
\begin{gather}
 \label{lambda-14}
 \lambda_+=\max\{|C_1|^2,|C_4|^2\},\;\lambda_-=\min\{|C_1|^2,|C_4|^2\},\\
 \label{Conc-14}
 C_g=2|C_1|\times|C_4|=|\sin 2\theta|,\\
 \label{P-14}
 P=\lambda_+-\lambda_-=\left||C_1|^2-|C_4|^2\right|=|\cos 2\theta|.
\end{gather}
The functions $C_g(\theta)$ and $P_g(\theta)$ are shown in Fig. \ref{Fig2}.

\noindent Maximally entangled unpolarized state ($C_g=1,\,P=0$) occurs when $|C_1|=|C_4|$ or $\theta=\pi/4$. Oppositely, TPPQ (\ref{case-a}) is disentangled and maximally polarized ($C_g=0,\,P=1$) if $|C_4|=0$ or $|C_1|=0$, i.e., if $\theta=0$ or $\theta=\pi/2$.

As follows from the general expression (\ref{Stokes-gen}) for the Stokes vector (per one photon) of TPPQ, in the case $C_2=C_3=0$ the vector ${\vec S}$ has only one non-zero component: $S_1=S_2=0$ and $S_3=|C_1|^2-|C_4|^2=\cos 2\theta$. In the Poincar\'{e} sphere  the vector ${\vec S}$ is directed along the $(V,H)$ axis and is given by 1/3 of the algebraic sum of all one-photon Stokes vectors of all photons presented in the the state (\ref{Stokes-gen}) with weighting factors $|C_1|^2$ and $|C_4|^2$, correspondingly, for horizontally and vertically polarized photons.

All these features of the three-photon state (\ref{case-a}) are practically identical to those of the biphoton qutrits of a special form, $C_1\ket{2_H}+C_4\ket{2_V}$ \cite{NJP,Che}. This direct analogy between the three-photon and biphoton states does not occur in other configurations of three-photon states, and one example of such configurations is considered in the following subsection.

\subsection{$C_1=C_4=0$}
\begin{figure}[h]
\includegraphics[width=6cm]{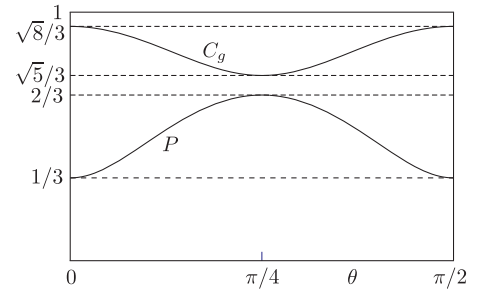}
\caption{{\protect\footnotesize {Degrees of entanglement $C_g$ and polarization $P$ (per one photon) as functions of the parameter $\theta$ of Eq. (\ref{qqrt-2-3})}}}\label{Fig3}
\end{figure}

This special case of a three-photon state is determined by the state vector of the form
\begin{gather}
 \nonumber
 \ket{\Psi_{2-3}}=C_2\ket{2_H,1_V}+C_3\ket{1_H,2_V}=\\
 \nonumber
 \cos\theta\ket{2_H,1_V}+e^{i\varphi}\sin\theta\ket{1_H,2_V}=\\
 \label{qqrt-2-3}
 \frac{1}{\sqrt{2}}\left(\cos\theta\,a_H^{\dag\,2}a_V^\dag+e^{i\varphi}\sin\theta\,a_H^\dag a_V^{\dag\,2}\right)\ket{0},
\end{gather}
where $|\theta|\leq\pi/2$, which corresponds to the constant $C_2$ taken real and positive, i.e., having a zero phase.  In accordance with Eqs. (\ref{Conc}) and (\ref{P-C-K}), the generalized concurrence and degree of polarization of the state (\ref{qqrt-2-3}) are given by
\begin{gather}
 \label{Cg-2-3}
 C_g=\frac{1}{3}\sqrt{8-12|C_2|^2|C_3|^2}=\frac{\sqrt{5+3\cos^22\theta}}{3},\\
 \label{P-2-3}
 P=\frac{1}{3}\sqrt{1+12|C_2|^2|C_3|^2}=\frac{\sqrt{4-3\cos^22\theta}}{3}.
\end{gather}
The variation ranges of these parameters are
\begin{equation}
 \label{ranges-2-3}
 \frac{\sqrt{5}}{3}\leq C_g\leq\frac{2\sqrt{2}}{3}\quad{\rm and}\quad\frac{2}{3}\geq P\geq\frac{1}{3}.
\end{equation}
Within these ranges, entanglement is maximal and degree of polarization is minimal at $\theta=0$ or $\theta=\pi/2$, i.e., at $C_3=0$ or $C_2=0$. And, oppositely, entanglement is minimal and degree of polarization is maximal at $\theta=\pi/4$ when $|C_2|=|C_3|=1/\sqrt{2}$. In other words, the single states $\ket{2_H,1_V}$ and $\ket{1_H,2_V}$ are more entangled and less polarized than their superpositions. This behavior is somewhat unexpected and contrasts with that of the states (\ref{case-a}) considered in the previous subsection. For the states (\ref{qqrt-2-3}) the dependencies of the generalized concurrence and the degree of polarization on the parameter $\theta$ are shown in Fig. \ref{Fig3}.

\begin{figure}[h]
\includegraphics[width=7cm]{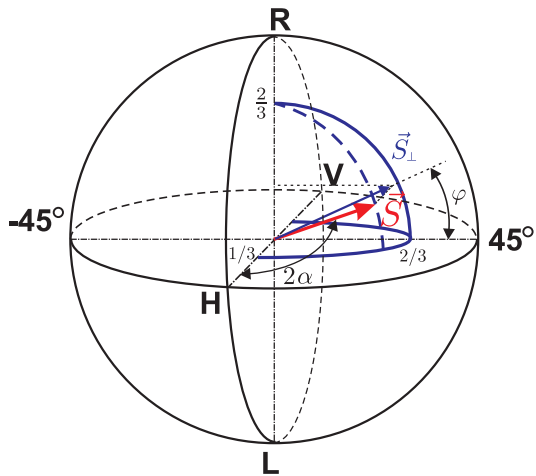}
\caption{{\protect\footnotesize {The Stokes vector of the state (\ref{qqrt-2-3}) per one photon.}}}\label{Fig4}
\end{figure}

The polarization Stokes vector of the state $\Psi_{2-3}$ is determined by the general expression (\ref{Stokes-gen}) with $C_1=C_4=0$
\begin{equation}
 \label{Stokes-2-3}
 {\vec S}=\frac{1}{3}\left(
 \begin{matrix}
 2\sin(2\theta)\cos\varphi\\
 2\sin(2\theta)\sin\varphi\\
 \cos(2\theta)
 \end{matrix}
 \right).
\end{equation}
Orientation of this Stokes vector in the Poincar\'{e} sphere is illustrated by Fig. \ref{Fig4}.
The solid blue lines in the horizontal and vertical planes determine the ending positions of the Stokes vector ${\vec S}$ in two cases: $\varphi=0$, $0\leq\theta\leq\pi/2$ and $\theta=\pi/4$, $0\leq\varphi\leq\pi/2$, whereas the dashed blue line in the vertical plane corresponds to the general case, $\varphi\neq 0$ and $\theta\neq\pi/4$.

In the case $\theta=0$ Eq. (\ref{Stokes-2-3}) yields: $S_1=S_2=0$ and $S_3=1/3$. This means that in this case the TPPQ Stokes vector ${\vec S}$ is directed along the $(H,V)$ axis, and its length equals $1/3$, which has a very simple explanation. At  $\theta=0$ the state (\ref{qqrt-2-3}) turns into the single basic state $\Psi_{2_H,1_V}$. In this state the lengths of collinear Stokes vectors of individual photons are equal 1 for horizontally and $-1$ for vertically polarized photons. The sum of these three individual Stokes vectors equals $1+1-1=1$. This is the length of the total Stokes vector of the state $\Psi_{2_H,1_V}$ as a whole. The Stokes vector per one photon is obtained from the total Stokes vector by means of division by the amount of photons, which gives $1/3$.

 If $\theta\neq 0$  the angle between the Stokes vector ${\vec S}$ (\ref{Stokes-2-3}) and the $(H,V)$ axis equals $2\alpha$, as shown in Fig. \ref{Fig4}, with the angle $2\alpha$ defined by the equation
\begin{equation}
 \label{2alpha}
 \cos 2\alpha=\left[\frac{\cos2\theta}{\sqrt{4-3\cos^22\theta}}\right].
\end{equation}
The second parameter of Eq. (\ref{qqrt-2-3}), $\varphi$, determines in this case the angle between the $(-45^\circ,45^\circ)$ axis and projection of the Stokes vector ${\vec S}$ on the vertical pane perpendicular to the $(H,V)$ axis. In a special case $\theta=\pi/4$, $\varphi=0$ the total Stokes vector (\ref{Stokes-2-3}) of the state (\ref{qqrt-2-3}) is directed along the axis $(-45^\circ, 45^\circ)$ in the Poincar\'{e} sphere, and its length equals 2/3.

\begin{figure}[h]
\includegraphics[width=7cm]{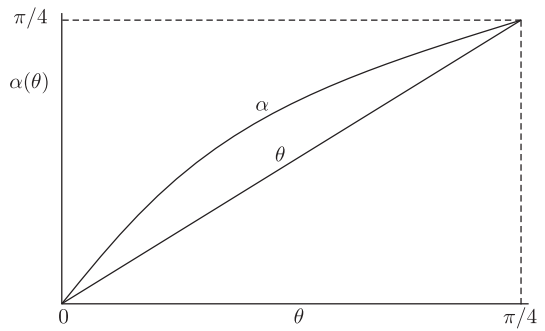}
\caption{{\protect\footnotesize {The function $\alpha(\theta)$ defined by Eq. (\ref{2alpha})}}}\label{Fig5}
\end{figure}

\noindent Only in the cases $\theta=0$ (or $\pi/2$) and $\theta=\pi/4$ Eq.(\ref{2alpha}) yields $\alpha=\theta$. In all other cases the angle $\alpha$ slightly exceeds $\theta$ as shown in Fig. \ref{Fig5}. Some explanation of these results are given below in the section 7 on the Schmidt modes and decomposition of the state (\ref{qqrt-2-3}).

\begin{figure}
\includegraphics[width=7cm]{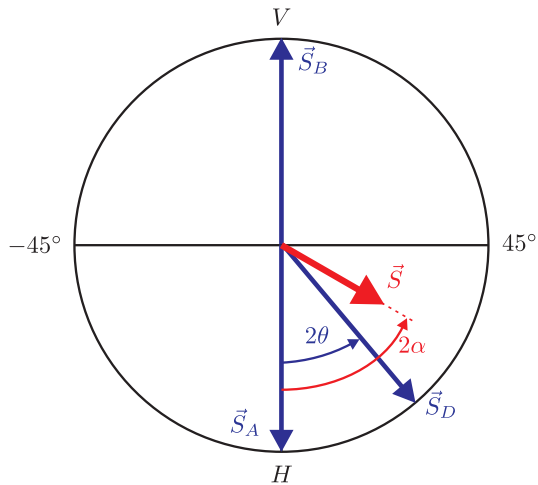}
\caption{{\protect\footnotesize {Horizontal plane of the Poincar\'{e} sphere. The TPPQ Stokes vector (red) and Stokes vectors of one-photon states generated by factorizing creation operators $A^\dag$, $B^\dag$, $D^\dag$ (blue) for the state (\ref{qqrt-2-3}) with $\varphi=0$.}}}\label{Fig6}
\end{figure}

The difference between $\theta$ and $\alpha$ is rather important for emphasizing the difference between the Stokes vectors of three- and two-photon states. As known \cite{Burl,MCh,Che}, the state vectors of biphoton states can be presented in the form $\ket{\Psi_{biph}}=NA^\dag B^\dag\ket{0}$, where $N$ is the normalizing factor and $A^\dag$ and $B^\dag$ are the single-photon creation operators, factorizing the biphoton state vector (a simple way of finding $A^\dag$ and $B^\dag$ and analysis of their features are given in Ref. \cite{Che}). The one-photon states $A^\dag\ket{0}$ and $B^\dag\ket{0}$ generated by these operators are characterized by their Stokes vectors ${\vec S}_A$ and ${\vec S}_B$. As known \cite{Burl,MCh,Che}, the biphoton Stokes vector ${\vec S}_{biph}$ is always located in the plane $\left\{{\vec S}_A,\,{\vec S}_B\right\}$ and is directed along the bisector of the angle between ${\vec S}_A$ and ${\vec S}_B$, i.e.,
${\vec S}_{biph}\|\left({\vec S}_A+{\vec S}_B\right)$. In the case of TPPQ, their state vectors can be shown to be representable in a similar form of a product of three one-photon creation operators $\ket{\Psi_{TPPQ}}=NA^\dag B^\dag D^\dag\ket{0}$ with $A^\dag$, $B^\dag$, and $D^\dag$ to be found in a way similar to that described in Ref. \cite{Che} for biphoton states. Then, it might be natural to think that the TPPQ Stokes vector ${\vec S}$ is parallel to the sum of three one-photon Stokes vectors ${\vec S}_A+{\vec S}_B+{\vec S}_D$. But in a general case this assumption appears to be wrong. This is clearly seen in the example of the state (\ref{qqrt-2-3}) we consider here. For this state the factorizing operators are evident: $A^\dag=a_H^\dag$, $B^\dag=a_V^\dag$, and $D^\dag=\frac{1}{\sqrt{2}}\left(\sin\theta\,a_H^\dag+e^{i\varphi}\sin\theta\,a_V^\dag\right)$. The Stokes vectors ${\vec S}_A$ and ${\vec S}_B$ are directed along the $(H,V)$ axis of the Poincar\'{e} sphere, they have equal absolute values ($=1$) but are oppositely directed. They cancel each other in the sum of three Stokes vectors ${\vec S}_{A,B,D}$ and, hence, ${\vec S}_A+{\vec S}_B+{\vec S}_D={\vec S}_D$. In the case $\varphi=0$ both ${\vec S}_D$ and TPPQ Stokes vector ${\vec S}$ are located in the horizontal plane of the Poincar\'{e} sphere, but they are not parallel to each other, as shown in Fig. \ref{Fig6}. The angles between these vectors and the $(H,V)$ axis are equal to $2\theta$ and $2\alpha$, correspondingly, for ${\vec S}_D$ and ${\vec S}$.
Thus, this example shows clearly that in a general case the TPPQ Stokes vector  is not parallel to the vectorial sum of one-photon Stokes vectors ${\vec S}_{A,B,D}$, and the assumed simple analogy with biphotons does not work for three-photon states. An alternative interpretation and other results arise in the approach based on the Schmidt decompositions, Schmidt modes and their Stokes vectors (see sections 6 and 7 below).

\subsection{Geometrical representation}

A rather interesting and picturesque geometrical way for characterizing the degree of entanglement is related to the use of the barycentric or trilinear coordinates of points in triangles \cite{BCC}. This method is applicable to special classes of TPPQ in which one of constants $C_1$, $C_2$, $C_3$, or $C_4$ equals zero and three remaining constants are real and positive. As shown in the inset of Fig. {\ref{Fig7}, for any point $O$ inside a triangle $ABC$ the sum of areas of smaller triangles $AOB$, $BOC$, and $AOC$ does not depend of the position of the point $O$ and equals the area of the triangle $ABC$, $S_{AOB}+S_{BOC}+S_{AOC}=S_{ABC}$. Owing to this condition one can identify positions of points inside the triangle $ABC$ with the TPPQ, and relative areas of smaller triangles with squared values of three nonzero TPPQ constants. E.g. as
$$C_1^2=\frac{S_{AOB}}{S_{ABC}},\,C_2^2=\frac{S_{BOC}}{S_{ABC}},\,C_3^2=\frac{S_{AOC}}{S_{ABC}}$$  in the case $C_4=0$. If the triangle $ABC$ is taken  equilateral, the TPPQ constants $C_i^2$ can be expressed in terms of distances $h_{1,2,3}$ from the point $O$ to the triangle sides $AB$, $BC$ and $AC$: $C_i^2=h_i/\sum_ih_i$. Characterization of position of points in triangles by their distances from the triangle sides corresponds to the definition of trilinear coordinates of these points, which are a special case of barycentric coordinates. For any given values of $h_i$ we find constants $C_i$ and, with the help of Eqs. (\ref{lambda-pm})-(\ref{entr}), a value of the reduced-state entropy $S_r=S_{rr}$. Then we color in different colors (from red to blue) regions corresponding to higher or lower levels of entanglement. The states indicated at the triangle apexes correspond to areas of small triangles opposite to these apexes and to distances from points inside the triangle to its sides opposite to apexes. In addition to the picture of Fig. \ref{Fig7} and in a similar way we can construct three other pictures corresponding to cases $C_1=0$, $C_2=0$, or $C_3=0$. Combined together, all these four pictures form a tetrahedron, one side of which is just the triangle shown in Fig. \ref{Fig7}.

\begin{figure}
\includegraphics[width=7cm]{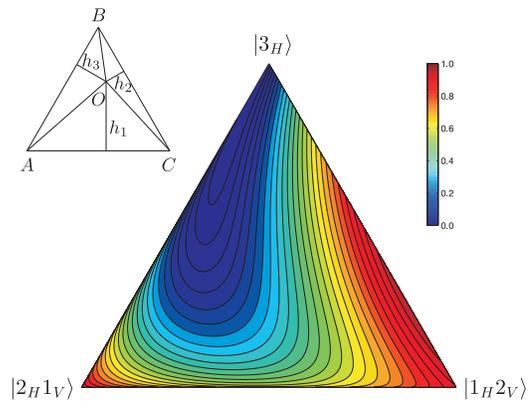}
\caption{{\protect\footnotesize {Entropy of the reduced states characterizing the degree of entanglement of TPPQ with $C_4=0$ and real and positive constants $C_{1,2,3}$.}}}\label{Fig7}
\end{figure}

\section{``Ideal$"$ Schmidt decomposition}

Schmidt decompositions are well defined for any pure bipartite states. In a general case, they present the decompositions of either wave functions or reduced density matrices of states in series of products of Schmidt modes $\psi_n$ and $\chi_n$
\begin{equation}
 \label{Schm-dec}
 \begin{matrix}
 \Psi(x_1,x_2) =\sum_n\sqrt{\lambda_n}\psi_n(x_1)\chi_n(x_2),\\\,\\\rho_r=\sum_n\lambda_n\ket{\psi_n}\bra{\chi_n},
 \end{matrix}
\end{equation}
where the Schmidt modes are defined for any given state as eigenfunctions of the reduced density matrix and $\lambda_n$ as its eigenvalues. The bases of Schmidt modes $\{\psi_n\}$ and $\{\chi_n\}$ are complete and in each of these two bases Schmidt modes are orthogonal to each other. The Schmidt decompositions are unique for any bipartite state as the only decompositions with single rather than double summation over numbers of modes. One of the main features of the Schmidt modes is that two particles of any given bipartite pair can appear only in adjoint Schmidt single-particle modes $\psi_n$ and $\chi_n$ (with the same number $n$), and never in modes with different numbers, $\psi_n$ and $\chi_{n^\prime}$. This makes Schmidt decompositions very appropriate, e.g., for characterization of entanglement, as well as for applications. In the case of biphoton states, because of symmetry of biphoton wave functions, the adjoint Schmidt modes coincide with each other, $\psi_n=\chi_n$, and the Schmidt decomposition of the wave function can be rewritten as the decomposition of the state vector \cite{Che}
\begin{equation}
 \label{Schm-dec-st-v}
 \ket{\Psi}=\sum_n\sqrt{\frac{\lambda_n}{2}}\,a_n^{\dag\,2}\ket{0},
\end{equation}
where $a_n^\dag$ are creation operators for photons in Schmidt modes. In the case of biphoton polarization qutrits there are only two polarization Schmidt modes, and the corresponding creation operators and eigenvalues of the reduced density matrix are denoted as $a_\pm^\dag$ and $\lambda_\pm$.

There is a class of TPPQ, the Schmidt decomposition of which most closely reminds the Schmidt decomposition of biphoton states (\ref{Schm-dec}), (\ref{Schm-dec-st-v}). This class of TPPQ is determined by the assumption that their state vectors can be reduced to the form
\begin{equation}
 \label{Schm-type-I}
 \ket{\Psi}=\frac{1}{\sqrt{6}}\left(\sqrt{\lambda_+}\;a_+^{\dag\,3}+\sqrt{\lambda_-}\;a_-^{\dag\,3}\right)\ket{0},
\end{equation}
where $a_+^\dag$ and $a_-^\dag$ are creation operators of photons in the orthogonal $+$ and $-$ Schmidt modes.
The simplest example is the TPPQ  (\ref{qqrt-st-vect}) with $C_2=C_3=0$
\begin{equation}
 \label{Schm-HV}
 \ket{\Psi}=\frac{1}{\sqrt{6}}\left(C_1a_H^{\dag\,3}+C_4a_V^{\dag\,3}\right)\ket{0}
\end{equation}
and with the wave function given by
\small
\begin{equation}
 \label{Schm-wf-HV}
 \Psi=C_1\left(1\atop 0\right)_1\left(1\atop 0\right)_2\left(1\atop 0\right)_3+
 C_4\left(0\atop 1\right)_1\left(0\atop 1\right)_2\left(0\atop 1\right)_3.
\end{equation}
\normalsize
In this case $\lambda_+=|C_1|^2$, $\lambda_-=|C_4|^2$, $a_+^\dag=a_H^\dag$ and $a_-^\dag=a_V^\dag$ and the Schmidt modes are $\psi_H=\left(1\atop 0\right)$ and $\psi_V=\left(0\atop 1\right)$. Any state of the form (\ref{Schm-type-I}) can be transformed to the form of Eq. (\ref{Schm-HV}) with the help of transformations equivalent to rotations of the Poincar\'{e} sphere and transforming the orthogonal Schmidt modes $\psi_+$ and $\psi_-$ to one-photon states with, correspondingly, horizontal and vertical polarizations, $\psi_H$ and $\psi_V$. Experimentally, such transformations are provided by appropriately installed quarter- and half-wavelength plates on a way of a three-photon beam.

On the other hand, the problem of choosing appropriate transformations can be formulated differently. Let us assume that originally we have a TPPQ of a general form (\ref{qqrt-st-vect}) with unknown coefficients  $C_{1,2,3,4}\neq 0$. The question is whether it's possible to transform it to the Schmidt-decomposition form (\ref{Schm-HV}) and under which conditions? In a general case, analytically, the discussed transformations are provided by the basis transformation formulas
\begin{gather}
 \label{transformation}
 \begin{matrix}
 a_+^\dag=a^\dag_H\cos\vartheta+\sin\vartheta e^{i\phi}a_V^\dag,\\
 a_-^\dag=-a^\dag_H\sin\vartheta+\cos\theta e^{i\phi}a_V^\dag;
 \end{matrix}\\
 \label{inverse}
 \begin{matrix}
 a_H^\dag=a^\dag_+\cos\vartheta-\sin\vartheta a_-^\dag,\\
 a_V^\dag=e^{-i\phi}\left(a^\dag_+\sin\vartheta+\cos\vartheta a_-^\dag\right),
 \end{matrix}
\end{gather}
where $\vartheta$ and $\phi$ are arbitrary real parameters of the transformation. Substitution of $a_H^\dag$ and $a_V^\dag$ of Eqs. (\ref{inverse}) into the general expression for the state vector of TPPQ (\ref{qqrt-st-vect}) reduces the latter to a similar form but with modified coefficients
\small
\begin{gather}
 \nonumber
 \ket{\Psi}=\\
 \label{qqrt-st-vect-pm}
 \left(\frac{{\widetilde C}_1}{\sqrt{6}}a_+^{\dag\,3}+\frac{{\widetilde C}_2}{\sqrt{2}}a_+^{\dag\,2}a_-^\dag+
 \frac{{\widetilde C}_3}{\sqrt{2}}a_+^\dag a_-^{\dag\,2}+ \frac{{\widetilde C}_4}{\sqrt{6}}a_-^{\dag\,3}\right)\ket{0}.
\end{gather}
\normalsize
If we want the transformed expression to have the form (\ref{Schm-type-I}), we have to require
\begin{equation}
 \label{condition-tilde}
 {\widetilde C}_2=0\;{\rm and}\; {\widetilde C}_3=0.
\end{equation}
Explicitly ${\widetilde C}_2$ and ${\widetilde C}_3$ are given by
\small
\begin{gather}
 \nonumber
 {\widetilde C}_2=-\sqrt{\frac{3}{2}}C_1\cos^2\vartheta\sin\vartheta
 +\frac{e^{-i\phi}}{\sqrt{2}}C_2(\cos^3\vartheta-2\cos\vartheta\sin^2\vartheta)\\
 \nonumber
 +\frac{e^{-2i\phi}}{\sqrt{2}}C_3(-\sin^3\vartheta+2\sin\vartheta\cos^2\vartheta)\\
 \label{C2-tilde}
 +\sqrt{\frac{3}{2}}e^{-3i\phi}C_4\sin^2\vartheta\cos\vartheta.
\end{gather}
\normalsize
and
\begin{gather}
 \nonumber
 {\widetilde C}_3=\sqrt{\frac{3}{2}}C_1\sin^2\vartheta\cos\vartheta
 +\frac{e^{-i\phi}}{\sqrt{2}}C_2(\sin^3\vartheta-2\sin\vartheta\cos^2\vartheta)\\
 \nonumber
 +\frac{e^{-2i\phi}}{\sqrt{2}}C_3(\cos^3\vartheta-2\cos\vartheta\sin^2\vartheta)\\
 \label{C3-tilde}
 +\sqrt{\frac{3}{2}}e^{-3i\phi}C_4\cos^2\vartheta\sin\vartheta.
\end{gather}
With real $\vartheta$ and $\phi$ equations (\ref{condition-tilde}) can be satisfied only if ${\widetilde C}_2$ (\ref{C2-tilde}) and ${\widetilde C}_3$ (\ref{C3-tilde}) are real. This condition puts limitations for phases $\varphi_{1,2,3,4}$ of the constants $C_{1,2,3,4}$. For example these conditions for phases can be taken in the form
\begin{equation}
 \label{phases}
 \varphi_1=0,\;\varphi_2=\phi,\;\varphi_3=2\phi,\;\varphi_4=3\phi.
\end{equation}
With these phases all phase factors in Eqs. (\ref{C2-tilde}) and (\ref{C3-tilde}) disappear and all constants $C_{1,2,3,4}$ are replaced by their absolute values. Then Eqs. (\ref{C2-tilde}) and (\ref{C3-tilde}) take the form of two cubic equations for $\tan\vartheta$. One solution of these equations is trivial, $\tan\vartheta=|C_2|=|C_3|=0$, which returns us to the state (\ref{case-a}). Two other solutions obey a couple of quadratic equations
\begin{equation}
 \label{quadratic}
 \begin{matrix}
 A\tan^2\vartheta+B\tan\theta+D=0,\\
 D\tan^2\vartheta-B\tan\theta+A=0,
 \end{matrix}
\end{equation}
where
\small
\begin{eqnarray}
 \label{ABD}
 \begin{matrix}
 A=|C_3|^2+|C_2|^2,\\
 B=\sqrt{3}\left(|C_1|\,|C_2|-|C_3|\,|C_4|\right),\\
 D=-2\left(|C_2|^2|C_3|^2\right)+
 \sqrt{3}\,\left(|C_2|\,|C_4|+|C_1|\,|C_3|\right).
 \end{matrix}
\end{eqnarray}
\normalsize
Two equations (\ref{quadratic}) are compatible only if $A=-D$, which yields
\begin{equation}
 \label{condition-C}
   |C_2|^2+|C_3|^2=\sqrt{3}\Big(|C_2|\,|C_4|+|C_3|\,|C_1|\Big).
\end{equation}
Eqs. (\ref{phases}) and (\ref{condition-C}) determine the complete set of conditions under which the TPPQ of a general form (\ref{qqrt-st-vect}) can be reduced to the form of the ``ideal$"$ Schmidt-decomposition  (\ref{Schm-type-I}). Under the same conditions two solutions of Eqs. (\ref{quadratic}) is given by
\begin{equation}
 \label{tangent}
   \tan\vartheta=\frac{-B\pm\sqrt{B^2+4A^2}}{2A}.
\end{equation}
This solutions together with Eqs. (\ref{transformation}), (\ref{phases}), (\ref{ABD}), and (\ref{condition-C}) can be used for finding explicitly the Schmidt-mode creation operators $a_+^\dag$ and $a_-^\dag$, as well as the Schmidt modes themselves $\ket{\psi_+}=a_+^\dag\ket{0}$ and $\ket{\psi_-}=a_-^\dag\ket{0}$.

Note that together with the normalization condition, the constraints (\ref{phases}) and (\ref{condition-C}) leave three free parameters: the phase $\phi=\varphi_2$ and two of four absolute values of the constants $C_{1,2,3,4}$, e.g., $|C_1|$ and $|C_2|$. Thus the described procedure defines a three-parametric manifold of states of TPPQ which can be reduced to the Schmidt-decomposition form of the type (\ref{Schm-type-I}).

\section{General form of the Schmidt decomposition}

Of course, there are many states of TPPQ which do not obey the conditions (\ref{phases}) and (\ref{condition-C}) and, thus, cannot be reduced to the form (\ref{Schm-type-I}). One example of such states is that of TPPQ (\ref{qqrt-st-vect}) with $C_1=C_4=0$. In this case the condition (\ref{condition-C}) is satisfied only if simultaneously $C_2=C_3=0$, which means no photons at all. However, the one-photon reduced density matrix $\rho_{rr}^{(1)}$ (\ref{rho-rr}) is known and well defined for all states of TPPQ, independently of any conditions for their parameters. Eigenfunctions of the reduced density matrix $\rho_{rr}^{(1)}$ are the one-photon Schmidt modes $\psi_\pm$. This means that one can write immediately the following general Schmidt decomposition for the  wave functions of arbitrary TPPQ wave functions of Eqs.  (\ref{qqrt-wf-gen}) and (\ref{Psi-3H})-(\ref{Psi-3V}):
\begin{equation}
 \label{Decomp-Psi-gen}
  \Psi(\sigma_1,\sigma_2,\sigma_3)=\sum_\pm\sqrt{\lambda_\pm}\;\psi_\pm(\sigma_1)\chi_\pm(\sigma_2,\sigma_3),
\end{equation}
where $\chi_\pm$ are the two-photon Schmidt modes, yet to be defined. For finding $\chi_\pm$ we can apply a procedure described in Ref. \cite{RocProc}, agreeing with the approach of the original work by E. Schmidt \cite{Schmidt} and based on the equations following directly from the decomposition (\ref{Decomp-Psi-gen})
\begin{equation}
 \label{Eq-for-chi}
 \sum_{\sigma_1}\psi_\pm^*(\sigma_1)\Psi(\sigma_1,\sigma_2,\sigma_3)=\sqrt{\lambda_\pm}\;\chi_\pm(\sigma_2,\sigma_3).
\end{equation}
As usual, the Schmidt modes are orthogonal and normalized
\begin{eqnarray}
 \nonumber
 \braket{\psi_\pm|\psi_\pm}=1,\,\braket{\psi_\mp|\psi_\pm}=0;\\
 \braket{\chi_\pm|\chi_\pm}=1,\,\braket{\chi_\mp|\chi_\pm}=0.
 \label{Schm-normalization}
\end{eqnarray}
Owing to this, the decomposition (\ref{Decomp-Psi-gen}) for the wave function yields immediately the standard decompositions for the reduced density matrices
\begin{equation}
 \label{Decomp-rho-gen}
 \rho_{rr}=\sum_\pm\lambda_\pm\ket{\psi_\pm}\bra{\psi_\pm},\;
 \rho_r=\sum_\pm\lambda_\pm\ket{\chi_\pm}\bra{\chi_\pm}.
\end{equation}
Owing to symmetry of the wave functions (\ref{Psi-3H})-(\ref{Psi-3V}), (\ref{qqrt-wf-gen}), the decomposition (\ref{Decomp-Psi-gen}) is also symmetric with respect to all transpositions of variables $\sigma_1,\,\sigma_2,\,\sigma_3$. But each single term in the sum over $\pm$ in (\ref{Decomp-Psi-gen}), separately from the other one, is not symmetric. Because of this, in a general case of arbitrary TPPQ the Schmidt decompositions for the wave function (\ref{Decomp-Psi-gen}) and density matrices (\ref{Decomp-rho-gen}) cannot be transformed to the decompositions of state vectors like those of Eqs. (\ref{Schm-dec-st-v}) and (\ref{Schm-type-I}). Moreover, in a general case each term in the sum over $\pm$ in Eq. (\ref{Decomp-Psi-gen}) cannot exist separately from the other one, and each single term in the decompositions (\ref{Decomp-Psi-gen}), (\ref{Decomp-rho-gen}) cannot represent any physically separable part of TPPQ. Nevertheless, the TPPQ Schmidt decompositions (\ref{Decomp-Psi-gen}) and (\ref{Decomp-rho-gen})  provide validity of the standard Schmidt-mode interpretation: if one of three TPPQ photons is found to be in one of two single-photon Schmidt modes, two other two photons of the same TPPQ triplet belong obligatory to the adjoint two-photon mode, $\chi_+$ for the single-photon mode $\psi_+$, and $\chi_-$ for the single-photon mode $\psi_-$, but not in any other combinations. Besides, the general-form Schmidt decomposition for the wave function (\ref{Decomp-Psi-gen}) appears to be very useful for finding the TPPQ Stokes vector and for their interpretation. By defining the TPPQ Stokes vector as the averaged vector of Pauli matrices ${\vec\sigma}^{\,(i)}$ acting in any $i$-th single-photon space of a polarization variable $\sigma_i$, $i=1,2,3$, we find from Eq. (\ref{Decomp-Psi-gen})
\begin{equation}
 \label{Stokes-gen-Schm-psi}
 {\vec S}=\braket{\Psi|{\vec\sigma}^{(1)}|\Psi}=\lambda_+{\vec S}_+^{\,\psi}+\lambda_-{\vec S}^{\,\psi}_-,
\end{equation}
where ${\vec S}_\pm^{\,\psi}=\braket{\psi_\pm|{\vec\sigma}^{(1)}|\psi_\pm}$ are the Stokes vectors of single-photon Shmidt modes $\psi_{\pm}$. On the other hand, if we take $i$=2 or 3, we get from Eq. (\ref{Decomp-Psi-gen}) an alternative representation for the same TPPQ Stokes vector
\begin{equation}
 \label{Stokes-gen-Schm-chi}
 {\vec S}=\braket{\Psi|{\vec\sigma^{(2)}}|\Psi}=\lambda_+{\vec S}^{\,\chi}_++\lambda_-{\vec S}^{\,\chi}_-,
\end{equation}
where ${\vec S}_\pm^{\,\chi}=\braket{\chi_\pm|{\vec\sigma}^{(2)}|\chi_\pm}$ are the Stokes vectors of the two-photon states $\ket{\chi_\pm}$.
Presentation of the TPPQ Stokes vector ${\vec S}$ in the form of sums of Schmidt-mode Stokes vectors ${\vec S}_\pm^{\,\psi}$ or ${\vec S}_\pm^{\,\chi}$ is analogous to similar results occurring in the case of biphoton polarization states (qutrits) \cite{Che}. Though, of course, if in the case of biphoton qutrits adjoint Schmidt modes are identical, in the case of TPPQ, evidently, $\psi_\pm\neq\chi_\pm$.

The Schmidt modes $\psi_+$ and $\psi_-$ represent pure one-photon polarization states, and they are orthogonal to each other. For these reasons their Stokes vectors ${\vec S}_+^{{\,\psi}}$ and ${\vec S}_-^{{\,\psi}}$ have a unit length each, $|{\vec S}_\pm^{{\,\psi}}|=1$, they are collinear with each other and with the TPPQ Stokes vector ${\vec S}$, and they are counter-directed, ${\vec S}_+^{{\,\psi}}=-{\vec S}_-^{{\,\psi}}$. As for the two-photon Schmidt modes $\chi_+$ and $\chi_-$ and their Stokes vectors ${\vec S}_+^{{\,\chi}}$ and ${\vec S}_-^{{\,\chi}}$, of course, the states $\ket{\chi_+}$ and $\ket{\chi_-}$ are orthogonal to each other. But the Stokes vectors ${\vec S}_+^{{\,\chi}}$ and ${\vec S}_-^{{\,\chi}}$ are determined by the reduced density matrices of the states $\ket{\chi_+}$ and $\ket{\chi_-}$ rather than by the states themselves. And, asw a rule, the reduced density matrices of these states, $\rho_{+\,r}$ and $\rho_{-\,r}$ are not orthogonal to each other, $\rho_{+\,r}\cdot\rho_{-\,r}\neq 0$. The only exception occurs in the case when both two-photon states $\ket{\chi_+}$ and $\ket{\chi_-}$ are disentangled, and this condition returns us to the case considered in the previous section and to states of the form (\ref{Schm-dec}), (\ref{Schm-dec-st-v}). If, however, each of the states  $\ket{\chi_+}$ and $\ket{\chi_-}$ is entangled, then inevitably $\rho_{+\,r}\cdot\rho_{-\,r}\neq 0$. Under these conditions the Stokes vectors ${\vec S}_+^{{\,\chi}}$ and ${\vec S}_-^{{\,\chi}}$ have lengths smaller than unit and they may be non-collinear with respect to each other and with respect to the TPPQ Stokes vector. But in any case, in accordance with Eq. (\ref{Stokes-gen-Schm-chi}) the sum of the Schmidt-mode Stokes vectors ${\vec S}_+^{{\,\chi}}$ and ${\vec S}_-^{{\,\chi}}$ with the weighting coefficients $\lambda_+$ and $\lambda_-$ gives the Stokes vector  of the TPPQ as a whole.

Below the described general features of the TPPQ Schmidt decompositions are specified in the consideration of TPPQ (\ref{qqrt-wf-gen}) with $C_1=C_4=0$.

\section{Schmidt modes and Stokes vectors of
TPPQ with $C_1=C_4=0$}
Thus, we consider now the state (\ref{qqrt-2-3}), i.e., the TPPQ with $C_1=C_4=0$, $C_2=\cos\theta$ and $C_3=\sin\theta e^{i\varphi}$, where $\theta$ and $\varphi$ are the only two independent constants characterizing this class of states. The wave function of the TPPQ under consideration is
\begin{equation}
 \label{wf-H2V-V2H}
 \Psi=\cos\theta\Psi_{2_H,1_V}+ e^{i\varphi}\sin\theta\Psi_{1_H,2_V}
\end{equation}
with $\Psi_{2_H,1_V}$ and $\Psi_{1_H,2_V}$ given by Eqs. (\ref{Psi-2H-1V}) and (\ref{Psi-1H-2V})

 For these parameters, the general expression (\ref{rho-rr}) for the the twice reduced TPPQ density matrix takes the form
\begin{equation}
 \rho_{rr}=
  \label{rho-rr-H2V-V2H}
 \frac{1}{6}\left(
 \begin{matrix}
  3+\cos 2\theta & 2e^{-i\varphi}\sin 2\theta\\
  2e^{i\varphi}\sin 2\theta & 3-\cos 2\theta
 \end{matrix}\right).
\end{equation}
As follows from Eqs. (\ref{lambda-pm}) and (\ref{Conc}), eigenvalues of $\rho_{rr}$ (\ref{rho-rr-H2V-V2H}) are given by
\begin{equation}
 \label{lambda-H2V-V2H}
 \lambda_\pm=\frac{1}{2}\left(1\pm\frac{1}{3}\sqrt{4-3\cos^2 2\theta}\right),
\end{equation}
whereas its generalized concurrence $C_g=2\sqrt{\lambda_+\lambda_-}$ and degree of polarization $P=\lambda_+-\lambda_-$ are given by Eqs. (\ref{Cg-2-3}) and (\ref{P-2-3}).

Eigenfunctions of the matrix $\rho_{rr}$ (\ref{rho-rr-H2V-V2H}) are easily found to be given by
\begin{equation}
 \label{Schm-modes-psi-H2V-V2H}
 \psi_+=\left(\cos\alpha\atop{e^{i\varphi}\sin\alpha }\right),\;\psi_-=\left(-\sin\alpha\atop{e^{i\varphi}\cos\alpha }\right),
\end{equation}
where $\alpha$ is determined by Eq. (\ref{2alpha}), and $\psi_\pm$ are the one-photon Schmidt modes of the decomposition (\ref{Decomp-Psi-gen}). At last, the two-photon Schmidt modes are determined from Eq. (\ref{Eq-for-chi}) with $\Psi$ and $\psi_\pm$ of Eqs. (\ref{wf-H2V-V2H}) and (\ref{Schm-modes-psi-H2V-V2H}). The results are given by
\begin{gather}
 \nonumber
 \chi_+=\frac{1}{\sqrt{3\lambda_+}}\Big(e^{-i\varphi}\sin\alpha\cos\theta\,\Psi_{2_H}
 +\cos(\alpha-\theta)\sqrt{2}\,\Psi_{1_H,1_V}\\
 \label{chi+H2V-V2H}
 +e^{i\varphi}\cos\alpha\sin\theta\,\Psi_{2_V}\Big)
\end{gather}
and
\begin{gather}
 \nonumber
 \chi_-=\frac{1}{\sqrt{3\lambda_-}}\Big(e^{-i\varphi}\cos\alpha\cos\theta\,\Psi_{2_H}
 -\sin(\alpha-\theta)\sqrt{2}\,\Psi_{1_H,1_V}\\
 \label{chi-H2V-V2H}
 -e^{i\varphi}\sin\alpha\sin\theta\,\Psi_{2_V}\Big).
\end{gather}

Eqs. (\ref{lambda-H2V-V2H})-(\ref{chi-H2V-V2H}) complete the definition of the Schmidt decomposition (\ref{Decomp-Psi-gen}) for the state (\ref{qqrt-2-3}), (\ref{wf-H2V-V2H}).
The next question concerns the Stokes vectors of this state. The total Stokes vector ${\vec S}$ of the state (\ref{qqrt-2-3}), (\ref{wf-H2V-V2H}) is given by Eq. (\ref{Stokes-2-3}). Its orientation in the Poincar\'{e} sphere is shown in Fig. \ref{Fig4}. It makes an angle $2\alpha$ with the $(H,V)$ axis. The Stokes vectors of the one-photon Schmidt modes $\psi_\pm$, ${\vec S}_\pm^{\,\psi}$, are easily found from the definition of these functions (\ref{Schm-modes-psi-H2V-V2H}) by means of constructing their density matrices and identifying them with the polarization matrices. The results are given by
\begin{gather}
 \nonumber
 S_{3\,\pm}^{\,\psi}=\pm\cos2\alpha,\;S_{1\,\pm}^{\,\psi}=\pm\sin2\alpha\cos\varphi,\\
 \label{Stokes-psi-H2V-H2V}
 S_{2\,\pm}^{\,\psi}=\pm\sin2\alpha\sin\varphi.
\end{gather}
Evidently, $\left|{\vec S}_\pm^{\,\psi}\right|=1$, ${\vec S}_+^{\,\psi}=-{\vec S}_-^{\,\psi}$, and both vectors ${\vec S}_+^{\,\psi}$ and ${\vec S}_-^{\,\psi}$ are located at the same axis as ${\vec S}$ in the Poincar\'{e} sphere.

As mentioned above, the Stokes vectors of biphoton states $\ket{\chi}_\pm$ are determined by their reduced density matrices. In fact, as these states are biphoton polarization qutrits, general expressions for their reduced density matrices are known \cite{NJP}. For arbitrary biphoton qutrits of the form $C_1\Psi_{2_H}+C_2\Psi_{1_H,1_V}+C_3\Psi_{2_V}$ the reduced density matrix is given by
\begin{equation}
 \nonumber
 \rho_r=\left(
 \begin{matrix}
 |C_1|^2+\frac{|C_2|^2}{2} & \frac{C_1C_2^*+C_2C_3^*}{\sqrt{2}}\\
 \frac{C_1^*C_2+C_2^*C_3}{\sqrt{2}} & |C_3|^2+\frac{|C_2|^2}{2}
 \end{matrix} \right).
\end{equation}
For the functions $\chi_+$ (\ref{chi+H2V-V2H}) and $\chi_-$ (\ref{chi-H2V-V2H}) this gives
\small
\begin{gather}
 \nonumber
 3\lambda_+\rho_{\chi_+\,r}=\\
 \label{rho-chi-r-+}
 \left(
 \begin{matrix}
  \sin^2\alpha\cos^2\theta+\cos^2(\alpha-\theta) & e^{-i\varphi}\cos(\alpha-\theta)\sin(\alpha+\theta)\\
  e^{i\varphi}\cos(\alpha-\theta)\sin(\alpha+\theta) & \cos^2\alpha\sin^2\theta+\cos^2(\alpha-\theta)
  \end{matrix} \right)
\end{gather}
\normalsize
and
\small
\begin{gather}
 \nonumber
 3\lambda_- \rho_{\chi_-\,r}=\\
 \label{rho-chi-r--}
 \left(
 \begin{matrix}
  \cos^2\alpha\cos^2\theta+\sin^2(\alpha-\theta) & -e^{-i\varphi}\cos(\alpha+\theta)\sin(\alpha-\theta)\\
  -e^{i\varphi}\cos(\alpha-\theta)\sin(\alpha+\theta) & \sin^2\alpha\cos^2\theta+\sin^2(\alpha-\theta)
  \end{matrix} \right)
\end{gather}
\normalsize
Found from here Stokes vectors of the Schmidt modes $\chi_+$ and $\chi_-$ are given by
\begin{equation}
 \label{S+chi}
 {\vec S}_+^{\,\chi}=\frac{1}{3\lambda_+}\left(
 \begin{matrix}
 2\cos\varphi\cos(\alpha-\theta)\sin(\alpha+\theta)\\
 2\sin\varphi\cos(\alpha-\theta)\sin(\alpha+\theta)\\
 \sin(\alpha-\theta)\sin(\alpha+\theta)
 \end{matrix}
 \right)
\end{equation}
and
\begin{equation}
 \label{S-chi}
 {\vec S}_-^{\,\chi}=\frac{1}{3\lambda_-}\left(
 \begin{matrix}
 -2\cos\varphi\sin(\alpha-\theta)\cos(\alpha+\theta)\\
 -2\sin\varphi\sin(\alpha-\theta)\cos(\alpha+\theta)\\
 \cos(\alpha-\theta)\cos(\alpha+\theta)
 \end{matrix}
 \right)
\end{equation}
The lengths of these vectors are equal to the degrees of polarization of the states $\ket{\chi_+}$ and $\ket{\chi_-}$
\begin{gather}
  \label{P-chi+pm}
  P_+=|{\vec S}_+^{\,\chi}|=\frac{\sin(\alpha+\theta)}{3\lambda_+}\sqrt{3\cos^2(\alpha-\theta)+1},\\
  \label{P-chi-pm}
  P_-=|{\vec S}_-^{\,\chi}|=\frac{\cos(\alpha+\theta)}{3\lambda_-}\sqrt{3\sin^2(\alpha-\theta)+1}
\end{gather}
Cosines of angles between the vectors ${\vec S}_+^{\,\chi}$, ${\vec S}_-^{\,\chi}$ and the $(H,V)$ axis are
\begin{gather}
 \label{angle-chi+}
  \cos\vartheta_+=\displaystyle\frac{S_{+\,3}^{\,\chi}}{|{\vec S}_+^{\,\chi}|}=\frac{\sin(\alpha-\theta)}{\sqrt{3\cos^2(\alpha-\theta)+1}},\\
 \label{angle-chi-}
 \cos\vartheta_-=\displaystyle\frac{S_{-\,3}^{\,\chi}}{|{\vec S}_-^{\,\chi}|}=\frac{\cos(\alpha-\theta)}{\sqrt{3\sin^2(\alpha-\theta)+1}}.
\end{gather}
These results show that, as expected, the Stokes vectors of the two photon Schmidt modes are not collinear neither to the Stokes vectors of the one-photon Schmidt modes nor to the Stokes vector of the  TPPQ as a whole. As for the latter, in terms of the reduced density matrices of the two-photon Schmidt modes (\ref{rho-chi-r-+}) and (\ref{rho-chi-r--}), the Stokes vector ${\vec S}$ is determined by the sum of $\rho_{\chi_+\,r}$ and $\rho_{\chi_-\,r}$ with weighting coefficients $\lambda_+$ and $\lambda_-$
\begin{equation}
 \label{sum of rho}
 \lambda_+\rho_{\chi_+\,r}+\lambda_+\rho_{\chi_-\,r}=\frac{1}{3}\left(
 \begin{matrix}
 1+\cos^2\theta & e^{-i\varphi}\sin 2\theta\\
 e^{i\varphi}\sin 2\theta &1+\sin^2\theta
 \end{matrix}\right)
\end{equation}
The TPPQ Stokes vector ${\vec S}$ determined by this summed density matrix is given by Eq. (\ref{Stokes-2-3}), i.e., it belongs to the plane of two Stokes vectors of two-photon Schmidt modes $\chi_\pm$ and equals to the vectorial sum of their non-collinear Stokes vectors ${\vec S}_\pm^{(\chi )}$ with the weighting factors $\lambda_\pm$. These results are illustrated by the picture of Fig. \ref{Fig8}, where all the involved Stokes vectors are shown schematically for the state (\ref{qqrt-2-3}) with $\varphi=0$.

\begin{figure}
\includegraphics[width=7cm]{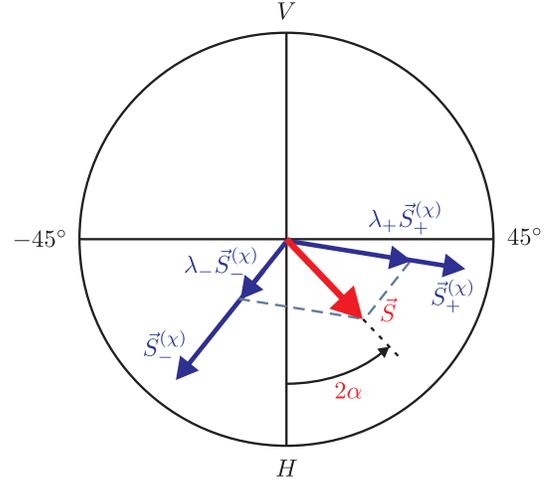}
\caption{{\protect\footnotesize {Horizontal plane of the Poincar\'{e}, the TPPQ Stokes vector (red) and Stokes vectors of the two-photon Schmidt modes for the states (\ref{qqrt-2-3}) with $\varphi=0$.}}}\label{Fig8}
\end{figure}

To conclude this section, let us show results following from the given above description in the simplest cases of $\theta=\alpha=0$ and $\theta=\alpha=\pi/4$, with $\varphi=0$.

1). $\theta=\alpha=\varphi=0$. In this case the state-vector of TPPQ consists of only one term, $\ket{\Psi}=\ket{2_H,1_V}$. In accordance with Eqs. (\ref{lambda-H2V-V2H})-(\ref{Stokes-psi-H2V-H2V}) and (\ref{S+chi})-(\ref{angle-chi-}) the main parameters of this state are: $\lambda_+=2/3$, $\lambda_-=1/3$, $P=1/3$, $\psi_+=\left(1\atop 0\right)$, $\psi_-=\left(0\atop 1\right)$, $\chi_+=\Psi_{1_H,1_V}$, and $\chi_-=\Psi_{2_H}$. The Stokes vector ${\vec S}_+^{\,\chi}$ equals zero, because the state $\ket{1_H,1_V}$ is maximally entangled and unpolarized \cite{NJP}. All other Schmidt-mode Stokes vectors have a unit length. All non-zero Schmidt-mode Stokes vectors, as well as the TPPQ Stokes vector ${\vec  S}$, are directed along the $(H,V)$ axis in the Poincar\'{e} sphere.

2). $\theta=\alpha=\pi/4,\,\varphi=0$. The state under consideration is $\frac{1}{\sqrt{2}}(\ket{2_H,1_V}+\ket{1_H,2_V})$, and  its parameters are: $\lambda_+=5/6$, $\lambda_-=1/6$, $P=2/3$, $\psi_+=\frac{1}{\sqrt{2}}\left(1\atop 1\right)$, $\psi_-=\frac{1}{\sqrt{2}}\left(-1\atop 1\right)$. The Stokes vectors ${\vec S}_+^{\,\psi}$ and ${\vec S}_-^{\,\psi}$ have unit lengths both, and they are oppositely directed along the $(-45^\circ,45^\circ)$ axis in the Poincar\'{e} sphere. The two-photon Schmidt modes are given by
\begin{gather}
 \nonumber
 \chi_+=\frac{1}{\sqrt{10}}\left(\Psi_{2_H}+2\sqrt{2}\,\Psi_{1_H,1_V}+\Psi_{2_V}\right),\\
 \label{chi-pi/4}
 \chi_-=\frac{1}{\sqrt{2}}\left(\Psi_{2_H}-\Psi_{2_V}\right).
\end{gather}
In this case the state $\ket{\chi_-}$ is maximally entangled and unpolarized, and its Stokes vector (\ref{S-chi}) equals zero. In the same time, the Stokes vector ${\vec S}_+^{\,\chi}$ has the only nonzero component $S_{+\,1}^{\,\chi}=\left|{\vec S}_+^{\,\chi}\right|=4/5$. The vector ${\vec S}_+^{\,\chi}$, as well as the total TPPQ Stokes vector ${\vec S}$, are directed in the positive direction of the $(-45^\circ,45^\circ)$ axis in the Poincar\'{e} sphere, and
\begin{equation}
 \label{S_P-pi/4}
 \left|{\vec S}\right|=P=\lambda_+S_{+\,1}^{\,\chi}=2/3.
\end{equation}

\section{Measurements}

If parameters of TPPQ $C_{1,2,3,4}$ are not known, an important problem is a possibility of their measuring. In principle, this can be done in  a way similar to that known for biphoton polarization qutrits \cite{NJP} and based on the use of a series of coincidence measurements. However, there is rather important difference related to the amount of photons in TPPQ and in biphoton states. As TPPQ states are states of three photons, the usual pair-coincidence measurements are insufficient and they have to be replaced by triple-coincidence measurements, when one registers only signals  coming simultaneously to a computer from three single-photon channels. A scheme of such measurements is shown in Fig. \ref{Fig9}.
\begin{figure}
\includegraphics[width=8cm]{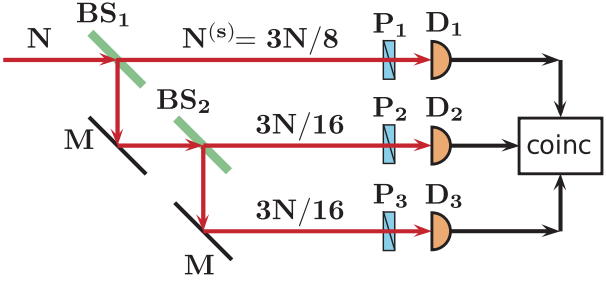}
\caption{{\protect\footnotesize {A scheme of triple coincidence measurements. BS - beamsplitters, M- mirrors, P-polarizers; N is the amount of photon triplets coming to the first beamsplitter during any given time $T$; $N^{(s)}$ shown in the upper and not shown in other channels indicates the amounts of single-photon states after beamsplitters.}}}\label{Fig9}
\end{figure}
 The scheme includes two nonselective beam splitters, a series of mirrors, three detectors and three polarizers. Each photon coming to a beamsplitter has equal 50$\%$  probabilities of transmission or reflection. As for triplets of photons  coming to a beam splitter, some of them pass or are reflected unsplit, and some others are split for one photon and pair of photons moving in different directions. By counting all possible combinations, we find amounts of single photons $N^{(s)}$ in all three channels for a given amount $N$ of triplets coming to the first beamslpitter: $N^{(s)}=\frac{3}{8}N$ after the first beamsplitter (in the upper channel in Fig. \ref{Fig9} and $N^{(s)}=\frac{3}{16}N$  after the second beamsplitter (in two lower channels). Only these photon triplets, completely split between three channels, give contributions to the triple-coincidence signals. Polarizers $(P_1,P_2,P_3)$ can be installed differently to permit transmission of differently polarized photons. In the first series of measurements it's sufficient to use only the following four combination polarizer installations:
\begin{equation}
 \label{polarizers}
 \left(\begin{matrix}P_1\\P_2\\P_3\end{matrix}\right)\rightarrow
 \left(\begin{matrix}H\\H\\H\end{matrix}\right),\;
 \left(\begin{matrix}H\\H\\V\end{matrix}\right),\;
 \left(\begin{matrix}V\\V\\H\end{matrix}\right),\;
 \left(\begin{matrix}V\\V\\V\end{matrix}\right).
 \end{equation}

 Note that there is a difference between the states $\ket{3_H}$ or $\ket{3_V}$ and $\ket{2_H,1_V}$ or $\ket{2_V,1_H}$. E.g., the first of these state can be registered with the installation of polarizers $HHH$, and every photon of each triplet can be registered in the lower channel. As for the state $\ket{2_H,1_V}$, it can be registered with the installation of polarizers $HHV$, and only one of three photons of each triplet can give contribution to the coincidence measurements. For this reason the amount of photons to be registered in the lower channel will be equal to $\frac{1}{16}$ rather than $\frac{3}{16}$, and the same is true also for the state $\ket{2_V,1_H}$. Finally, the probabilities of arising for states $\ket{3_H}$,  $\ket{2_H,1_V}$, $\ket{2_V,1_H}$, and $\ket{3_V}$ in TPPQ are equal to $|C_1|^2, |C_2|^2, |C_3|^2, |C_4|^2$. With all these comments taken into account we find the following relations between the absolute values of the TPPQ parameters  $C_{1,2,3,4}$ and amount of coincidence clicks of detectors in the schemes of Fig. \ref{Fig9} with the polarizer-installation schemes indicated in Eq. (\ref{polarizers}):
\begin{equation}
 \label{counts}
 \begin{matrix}[1.5]
 N_{HHH}=\frac{3}{16}\eta_1\eta_2\eta_3 N |C_1|^2,\\
 N_{HHV}=\frac{1}{16}\eta_1\eta_2\eta_3 N |C_2|^2,\\
 N_{VVH}=\frac{1}{16}\eta_1\eta_2\eta_3 N |C_3|^2,\\
 N_{VVV}=\frac{3}{16}\eta_1\eta_2\eta_3 N |C_4|^2,
 \end{matrix}
\end{equation}
where $\eta_1,\,\eta_2,\,\eta_3$ are efficiencies of tree detectors in the scheme of Fig. \ref{Fig9}. As the constants $C_{1,2,3,4}$ obey the normalization condition $|C_1|^2+|C_2|^2+|C_3|^2+|C_4|^2$, we can construct the sum of the measurement results, which does not depend on the constants $C_{1,2,3,4}$:
\begin{gather}
 \nonumber
 \Sigma=N_{HHH}+3N_{HHV}+3N_{VVH}+N_{VVV}\\
 \label{Sigma}
 =\frac{3\eta_1\eta_2\eta_3 N}{16},
\end{gather}
which gives
\begin{gather}
 \nonumber
 |C_1|^2=\frac{N_{HHH}}{\Sigma},\;|C_2|^2=\frac{3N_{HHV}}{\Sigma},\\
 \label{C squared via Sigma}
 |C_3|^2=\frac{3N_{VVH}}{\Sigma},\;|C_4|^2=\frac{N_{VVV}}{\Sigma}.
\end{gather}
Note that though both the direct results of measurements, $N_{HHH}$ etc. (\ref{counts}), and their sum $\Sigma$ (\ref{Sigma}) are proportional to the product of detector efficiencies $\eta_1\eta_2\eta_3$ and the the amount of initial photon triplets $N$, their ratios (\ref{C squared via Sigma}) are independent of these parameters.

The next step is measuring phases $\varphi_{1,2,3,4}$ of the constants $C_{1,2,3,4}$. Actually, as one of this phases can be taken equal zero, there are only three phases to be measured, e.g., $\varphi_2$, $\varphi_3$, and $\varphi_4$,  with $\varphi_1=0$ (if $|C_1|\neq 0$).  These phases can be found from a series of equations arising when the same measurements as described above are repeated with the polarizers turned for angles $45^\circ$  and $135^\circ$ (correspondingly, instead of the horizontal and vertical directions). The arising equation are identical to (\ref{counts})-(\ref{C squared via Sigma}) with substitutions
\begin{gather}
 \nonumber
 N_{HHH}\rightarrow N_{3_{45^\circ}},\; N_{HHV}\rightarrow N_{2_{45^\circ},1_{135^\circ}},\\
 \nonumber
 N_{VVH}\rightarrow N_{2_{135^\circ}, 1_{45^\circ}},\;N_{VVV}\rightarrow N_{3_{135^\circ}}
\end{gather}
and $C_i\rightarrow C_i^{45^\circ}$, where $i=1,2,3,4$ and $C_i^{45^\circ}$ are the same TPPQ parameters as in Eq. (\ref{qqrt-st-vect}) but in a basis turned for $45^\circ$:
\begin{gather}
 \nonumber
 C_1^{45^\circ}=\frac{1}{2\sqrt{2}}\left[C_1+C_4+\sqrt{3}(C_2+C_3)\right],\\
 \nonumber
 C_2^{45^\circ}=\frac{1}{2\sqrt{2}}\left[\sqrt{3}(-C_1+C_4)C_1+(-C_2+C_3)\right],\\
 \nonumber
 C_3^{45^\circ}=\frac{1}{2\sqrt{2}}\left[\sqrt{3}(C_1+C_4)C_1-(C_2+C_3)\right],\\
 \label{C-i-45}
 C_4^{45^\circ}=\frac{1}{2\sqrt{2}}\left[(-C_1+C_4)+\sqrt{3}(C_2-C_3)\right].
\end{gather}
The squared absolute values of these constants can be expressed via the measured amounts of triple-coincidence counts in the turned basis. These equalities will contain unknown phases in the form of superpositions of $\cos\varphi_i$ and $\cos(\varphi_i-\varphi_j)$ with $i,j=2,3,4$, which can be solved numerically. To be specific, let us show only one (first) of these equations
\begin{gather}
 \nonumber
 \frac{N_{3_{45^\circ}}}{\Sigma}=\left|C_1^{45^\circ}\right|^2=\frac{1}{8}\Big\{|C_1|^2+3|C_2|^2+3|C_3|^2+|C_4|^2+\\
 \nonumber
 2|C_1|\,|C_4|\cos\varphi_4 + 2\sqrt{3}\,|C_1|(|C_2|\cos\varphi_2+|C_3|\cos\varphi_3)+\\
 \nonumber
 2\sqrt{3}\,|C_4|\left[|C_2|\cos(\varphi_2-\varphi_4)+|C_3|\cos(\varphi_3-\varphi_4)\right]+\\
 \label{turned basis}
 6|C_2||C_3|\cos(\varphi_2-\varphi_3)\Big\}.
 \end{gather}
 All other equations are similar and they differ from this one by coefficients and signs in front of terms in Eq. (\ref{turned basis}). Only tree of four equations of the type (turned basis) are independent from each other, and they are sufficient for finding three unknown phases, $\varphi_{2,3,4}$.

 A much simpler scheme of measurements can be used for finding parameters of TPPQ which can be reduced to the form (\ref{Schm-type-I}). Both this form and a scheme of its measurement (Fig. \ref{Fig10}) indicate a deep analogy between the TPPQ of the type (\ref{Schm-type-I}) and biphoton polarization qutrits \cite{NJP,Che}.

Then the beam can be sent to a polarizing beamsplitter. If the latter is installed in a standard way, it provides transmission of all horizontally polarized photons
\vskip 6pt
 \begin{figure}[h]
\includegraphics[width=7cm]{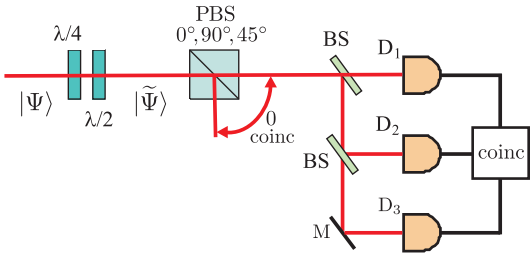}
\caption{{\protect\footnotesize {A scheme for measuring parameter of TPPQ of the form (\ref{Schm-type-I}).}}}\label{Fig10}
\end{figure}
At first, the expression for the TPPQ state vector (\ref{Schm-type-I}) can be further simplified with the help of properly installed $\lambda/4$- and $\lambda/2$-plates on a way of a three-photon beam to transform the Schmidt mode $\psi_+$ into $\psi_H$ and $\psi_-$ into $e^{i\phi}\psi_V$, where $\phi$ is some unknown phase. For the TPPQ state vector in the Schmidt-mode representation (\ref{Schm-type-I}) this gives
\begin{equation}
 \label{transform}
 \ket{\Psi}\rightarrow \ket{{\widetilde \Psi}}=\sqrt{\lambda_+}\,\ket{3_H}+e^{3i\phi}\sqrt{\lambda_-}\,\ket{3_V}.
\end{equation}
 An experimental criterion that the $\lambda/4$- and $\lambda/2$-plates are installed correctly is the zero coincidence signal between two channels immediately after PBS (any wrong installations do not provide the described transformation and give rise to a non-zero coincidence signal). After the transformation (\ref{transform}), as $\lambda_++\lambda_-=1$, the state $\ket{\widetilde \Psi}$ is characterized by two parameters only, e.g., $\lambda_+$ and the phase $\phi$. In accordance with what was proposed for biphoton qutrits \cite{Che}, for measuring  $\lambda_+$ and $\phi$, one can send the beam to the polarizing beamsplitter and then to the triple-coincidence scheme in one of the channels after PBS. The triple-coincidence scheme can be simplified compared to that shown in Fig. \ref{Fig9} because now polarizers are not needed. The amount of coincidence counts in the transmission channel of PBS installed in a standard way equals to
 \begin{equation}
  \label{N-0}
   N_{0^\circ}=\frac{3}{16}\eta_1\eta_2\eta_3N\lambda_+,
 \end{equation}
where $N$ is the amount of photon triplets coming to PBS per a given time $\Delta t$. If in the second series of measurements we turn PBS for $90^\circ$, the transmission channel will becomes open only for vertically polarized photons, and the amount of coincidence counts equals to
\begin{equation}
 \label{N-90}
 N_{90^\circ}=\frac{3}{16}\eta_1\eta_2\eta_3N\lambda_+.
\end{equation}
The sum of these two results is
\begin{equation}
 \label{N0+N90}
 \Sigma=N_++N_-=\frac{3}{16}\eta_1\eta_2\eta_3N.
\end{equation}
This sum does not depend of $\lambda_\pm$ and can be used for normalization to give expressions for $\lambda_+$ and $\lambda_-$ in terms of experimentally measurable amount of coincidence counts
\small
\begin{equation}
 \label{lambda via Sigma}
  \lambda_+=\frac{N_+}{\Sigma}=\frac{N_+}{N_++N_-},\; \lambda_-=\frac{N_-}{\Sigma}=\frac{N_-}{N_++N_-}.
\end{equation}
\normalsize
For measuring the phase $\phi$ PBS has to be turned for $45^\circ$. Then the transmission channel is open only for $45^\circ$-polarized photons,
and the state describing such photons is
\begin{equation}
 \label{psi-transm}
 \ket{\Psi}_{transm}=\frac{1}{2\sqrt{2}}\left(\sqrt{\lambda_+}+e^{3i\phi}\sqrt{\lambda_-}\right)\ket{3_{45^\circ}}.
\end{equation}
The corresponding amount of the triple-coincidence counts is
\begin{equation}
 \label{N-45}
  N_{45^\circ}=\frac{3}{8\times16}\eta_1\eta_2\eta_3N(1+C_g\cos{3\phi}),
\end{equation}
 where $C_g=2\sqrt{\lambda_+\lambda_-}$ is the generalized concurrence. The equation expressing the phase $\phi$ in terms of the experimentally measurable amounts of counts is given by
\begin{equation}
  \label{phi via Sigma}
  \frac{1}{8}\left(1+C_g\cos 3\phi\right)=\frac{N_{45^\circ}}{\Sigma}=\frac{N_{45^\circ}}{N_{0^\circ}+N_{90^\circ}}.
\end{equation}

\section{Conclusions}
Three-photon ququarts considered in this work are states of three photons with either coinciding or different polarizations but with identical given frequencies and propagation directions. In these, as well in any other, states photons are indistinguishable, owing to which the three-photon wave functions are symmetric with respect to any transpositions of three polarization variables of photons. The density matrix $\rho$ constructed from the TPPQ wave function is reduced with respect to either one or two polarization variables to give rise to once- and twice-reduced density matrices, $\rho_r$ and $\rho_{rr}$. Eigenvalues of these reduced density matrices are found in a general form (\ref{lambda-pm}). Owing to indistinguishability of photons and symmetry of the TPPQ wave functions, for any given TPPQ there is only one set of eigenvalues of the reduced density matrices, independent of which variables are chosen for taking traces of the full and once-reduced density matrices. As well as in the case of two-photon states (polarization qutrits), the reduced density matrices of TPPQ have only two non-zero eigenvalues. They are used to find in a general form the degree of entanglement of TPPQ, degree of polarization and Stokes vectors. The degree of entanglement can be characterized by such parameters as the generalized concurrence $C_g$ (\ref{Conc}), Schmidt parameter $K$ (\ref{K}), and von Neumann entropy of the reduced two-photon or one-photon states $S_r=S_{rr}$ (\ref{entr}). All three parameters are found to be absolutely compatible with each other. Analogously to biphoton qutrits, the minimal and maximal degrees of entanglement of TPPQ are characterized by  $C_{g\,\min}=S_{r\,\min}=0,\,K_{\min}=1$ and $C_{g\,\max}=S_{r\,\max}=1,\,K_{\max}=2$. Moreover, the generalized concurrence and the degree of polarization $P$ (per one photon) are found to be connected with each other by the same relation as in the case of biphotn qutrits \cite{NJP}, $C_g^2+P^2=1$. One-photon ($\psi_\pm$) and two-photon ($\chi_\pm$) eigenfunctions of the twice- and once-reduced density matrices provide a possibility of constructing Schmidt decompositions for wave functions and reduced density matrices of TPPQ with arbitrary parameters. A very special case, most closely reminding the case of biphoton qutrits, is that of the TPPQ consisting of a superposition of only two terms (\ref{Psi-3H}) and (\ref{Psi-3H}), with tree photons in each term having coinciding polarizations, either horizontal or vertical (\ref{case-a}). We found conditions (\ref{phases}), (\ref{condition-C}) under which the Schmidt decomposition is ``ideal$"$ , i.e. under which an arbitrary TPPQ can be reduced to the form (\ref{Schm-type-I}). This is a three-parametric class of TPPQ, and for such states the Schmidt decomposition occurs both for the wave functions and state vectors, as in the case of biphotons.

In a general case the Schmidt decomposition is found to occur for wave functions and reduced density matrices of all TPPQ but may be unachievable for state vectors. Nevertheless the Schmidt decomposition appears to be very useful for establishing connection with the polarization Stokes vectors. As shown, the polarization Stokes vector of an arbitrary TPPQ can be presented as the sum of Stokes vectors of Schmidt modes with the weighting factors equal to the eigenvalues of the reduced density matrices, (\ref{Stokes-gen-Schm-psi}), (\ref{Stokes-gen-Schm-chi}). This relation is analogous to that occurring for biphotons, and in the case of TPPQ it is found to occur for both one-photon and two-photon Schmidt modes. On the other hand, as well known \cite{Burl}, in the case of biphotons there is another interpretation of  Stokes vectors as proportional to the sum of Stokes vectors of two one-photon states generated by two creation operators factorizing biphoton state vectors. As shown above, this interpretation has no analogous counterpart in the case of TPPQ. TPPQ state vectors also can be reduced to the form with the product of three one-photon creation operators. These operators generate three one-photon states and three corresponding Stokes vectors. But, as shown, in a general case the vectorial sum of these three Stokes vectors is not related in any way to the TPPQ Stokes vector and, thus, the model working well for biphotons does not work in the case of TPPQ.

A series of schemes for experimental measurement of the TPPQ parameters is suggested. In principle, other questions not addressed in this work  may be related to possibilities of using TPPQ in practice. We hope to return to this research area elsewhere. But on the other hand, real applications will become possible only when a sufficiently easy and reliable way of producing three-photon states will be worked out, for which we hope to happen in a not too distant future.

\section*{Acknowledgement}
The work is supported partially by the grant RFBR 14-02-00811

\end{document}